\documentclass[12pt]{article}
\usepackage{amsmath}
\usepackage{graphicx}
\usepackage[square,sort,comma,numbers]{natbib}
\usepackage{url} 

\usepackage[nolists]{endfloat}

\newcommand{\blind}{0}

\usepackage{subfiles}

\usepackage{multirow}

\usepackage[top=3.2cm, bottom=3.2cm, left=2.4cm, right=2.4cm]{geometry}

\usepackage{fancyhdr}

\usepackage[bottom]{footmisc}

\usepackage{lineno}

\usepackage{hyperref}
\hypersetup{
    colorlinks=true,
    linkcolor=blue,
    citecolor=blue,
    filecolor=magenta,      
    urlcolor=blue,
}

\usepackage{array}
\newcolumntype{P}[1]{>{\centering\arraybackslash}p{#1}} 

\usepackage{siunitx}
\usepackage{dcolumn}
\newcolumntype{L}{D{.}{.}{2,2}}

\usepackage{bookmark}

\usepackage{bm}

\usepackage{setspace}

\usepackage{amsmath}
\usepackage{amsfonts}
\usepackage{amssymb}
\usepackage{amsthm}

\usepackage[T1]{fontenc}
\usepackage{mathtools}

\usepackage{mathptmx}

\usepackage{textcomp}

\usepackage{color, colortbl}
\usepackage[svgnames]{xcolor}

\usepackage{enumitem}

\usepackage{framed}

\usepackage{cancel}

\usepackage{amsmath}
\usepackage[linesnumbered,ruled]{algorithm2e}

\usepackage{graphicx}
\usepackage{caption}
\usepackage{subcaption}
\usepackage{float} 

\usepackage{chngcntr}
\usepackage{aliascnt}
\usepackage{tikz}
\usetikzlibrary{shapes.misc,calc}
\usepackage{tcolorbox}
\tcbuselibrary{skins,theorems,breakable}

\usepackage{epstopdf}

\let\emph\relax 
\DeclareTextFontCommand{\emph}{\bfseries}

\def\newshortcut#1#2{%
\let#1=\undefined
\newcommand{#1}{#2}}

\DeclareSymbolFont{usualmathcal}{OMS}{cmsy}{m}{n}
\DeclareSymbolFontAlphabet{\mathcal}{usualmathcal}
\DeclareSymbolFontAlphabet{\mathbb}{AMSb}

\newshortcut{\C}{\mathbb{C}}
\newshortcut{\F}{\mathbb{F}}
\newshortcut{\N}{\mathbb{N}}
\newshortcut{\Q}{\mathbb{Q}}
\newshortcut{\R}{\mathbb{R}}
\newshortcut{\T}{\mathbb{T}}
\newshortcut{\Z}{\mathbb{Z}}
\newshortcut{\D}{\mathbb{D}}

\newshortcut{\cA}{\mathcal{A}}
\newshortcut{\cB}{\mathcal{B}}
\newshortcut{\cC}{\mathcal{C}}
\newshortcut{\cD}{\mathcal{D}}
\newshortcut{\cE}{\mathcal{E}}
\newshortcut{\cF}{\mathcal{F}}
\newshortcut{\cG}{\mathcal{G}}
\newshortcut{\cH}{\mathcal{H}}
\newshortcut{\cI}{\mathcal{I}}
\newshortcut{\cJ}{\mathcal{J}}
\newshortcut{\cK}{\mathcal{K}}
\newshortcut{\cL}{\mathcal{L}}
\newshortcut{\cM}{\mathcal{M}}
\newshortcut{\cN}{\mathcal{N}}
\newshortcut{\cO}{\mathcal{O}}
\newshortcut{\cP}{\mathcal{P}}
\newshortcut{\cQ}{\mathcal{Q}}
\newshortcut{\cR}{\mathcal{R}}
\newshortcut{\cS}{\mathcal{S}}
\newshortcut{\cT}{\mathcal{T}}
\newshortcut{\cU}{\mathcal{U}}
\newshortcut{\cV}{\mathcal{V}}
\newshortcut{\cW}{\mathcal{W}}
\newshortcut{\cX}{\mathcal{X}}
\newshortcut{\cY}{\mathcal{Y}}
\newshortcut{\cZ}{\mathcal{Z}}

\newshortcut{\fA}{\mathfrak{A}}
\newshortcut{\fB}{\mathfrak{B}}
\newshortcut{\fC}{\mathfrak{C}}
\newshortcut{\fD}{\mathfrak{D}}
\newshortcut{\fE}{\mathfrak{E}}
\newshortcut{\fF}{\mathfrak{F}}
\newshortcut{\fG}{\mathfrak{G}}
\newshortcut{\fH}{\mathfrak{H}}
\newshortcut{\fI}{\mathfrak{I}}
\newshortcut{\fJ}{\mathfrak{J}}
\newshortcut{\fK}{\mathfrak{K}}
\newshortcut{\fL}{\mathfrak{L}}
\newshortcut{\fM}{\mathfrak{M}}
\newshortcut{\fN}{\mathfrak{N}}
\newshortcut{\fO}{\mathfrak{O}}
\newshortcut{\fP}{\mathfrak{P}}
\newshortcut{\fQ}{\mathfrak{Q}}
\newshortcut{\fR}{\mathfrak{R}}
\newshortcut{\fS}{\mathfrak{S}}
\newshortcut{\fT}{\mathfrak{T}}
\newshortcut{\fU}{\mathfrak{U}}
\newshortcut{\fV}{\mathfrak{V}}
\newshortcut{\fW}{\mathfrak{W}}
\newshortcut{\fX}{\mathfrak{X}}
\newshortcut{\fY}{\mathfrak{Y}}
\newshortcut{\fZ}{\mathfrak{Z}}
\newshortcut{\fa}{\mathfrak{a}}
\newshortcut{\fb}{\mathfrak{b}}
\newshortcut{\fc}{\mathfrak{c}}
\newshortcut{\fd}{\mathfrak{d}}
\newshortcut{\fe}{\mathfrak{e}}
\newshortcut{\ff}{\mathfrak{f}}
\newshortcut{\fg}{\mathfrak{g}}
\newshortcut{\fh}{\mathfrak{h}}
\newshortcut{\fj}{\mathfrak{j}}
\newshortcut{\fk}{\mathfrak{k}}
\newshortcut{\fl}{\mathfrak{l}}
\newshortcut{\fm}{\mathfrak{m}}
\newshortcut{\fn}{\mathfrak{n}}
\newshortcut{\fo}{\mathfrak{o}}
\newshortcut{\fp}{\mathfrak{p}}
\newshortcut{\fq}{\mathfrak{q}}
\newshortcut{\fr}{\mathfrak{r}}
\newshortcut{\fs}{\mathfrak{s}}
\newshortcut{\ft}{\mathfrak{t}}
\newshortcut{\fu}{\mathfrak{u}}
\newshortcut{\fv}{\mathfrak{v}}
\newshortcut{\fw}{\mathfrak{w}}
\newshortcut{\fx}{\mathfrak{x}}
\newshortcut{\fy}{\mathfrak{y}}
\newshortcut{\fz}{\mathfrak{z}}

\newshortcut{\vaprhi}{\varphi}

\newshortcut{\arsinh}{\operatorname{arsinh}}
\newshortcut{\arcosh}{\operatorname{arcosh}}
\newshortcut{\artanh}{\operatorname{artanh}}
\newshortcut{\arcsch}{\operatorname{arcsch}}
\newshortcut{\arsech}{\operatorname{arsech}}
\newshortcut{\arcoth}{\operatorname{arcoth}}
\newshortcut{\sign}{\operatorname{sign}}
\newshortcut{\sinc}{\operatorname{sinc}}
\newshortcut{\Log}{\operatorname{Log}}
\newshortcut{\Ei}{\operatorname{Ei}}
\newshortcut{\Si}{\operatorname{Si}}
\newshortcut{\Ci}{\operatorname{Ci}}

\usepackage{amsmath}

\newshortcut{\d}{\partial}
\newshortcut{\del}{\nabla}
\newshortcut{\grad}{\operatorname{grad}}
\newshortcut{\div}{\operatorname{div}}
\newshortcut{\Re}{\operatorname{Re}}
\newshortcut{\Im}{\operatorname{Im}}
\newshortcut{\PV}{\operatorname{PV}}
\newshortcut{\supp}{\operatorname{supp}}
\newshortcut{\spec}{\operatorname{spec}}
\newshortcut{\dist}{\operatorname{dist}}

\newshortcut{\bigplus}{\mathop{\Large+}}

\newshortcut{\bmu}{\bm{\mu}}
\newshortcut{\bnu}{\bm{\nu}}
\newshortcut{\bbeta}{\pmb{\beta}}
\newshortcut{\btheta}{\bm{\theta}}

\newcommand{\transpose}{^\prime}

\usepackage[utf8]{inputenc}
\usepackage{listings}
\usepackage{color}
 
\definecolor{codegreen}{rgb}{0,0.6,0}
\definecolor{codegray}{rgb}{0.5,0.5,0.5}
\definecolor{codepurple}{rgb}{0.58,0,0.82}
\definecolor{backcolour}{rgb}{0.95,0.95,0.92}
 
\lstdefinestyle{mystyle}{
    backgroundcolor=\color{backcolour},   
    commentstyle=\color{codegreen},
    keywordstyle=\color{magenta},
    numberstyle=\tiny\color{codegray},
    stringstyle=\color{codepurple},
    basicstyle=\footnotesize,
    breakatwhitespace=false,         
    breaklines=true,                 
    captionpos=b,                    
    keepspaces=true,                 
    numbers=left,                    
    numbersep=5pt,                  
    showspaces=false,                
    showstringspaces=false,
    showtabs=false,                  
    tabsize=2
}
 
\def\sinc{\mathop{\rm sinc}\nolimits}

\def\div{\mathop{\rm div}\nolimits}
\def\tr{\mathop{\rm tr}\nolimits}

\def\dist{\mathop{\rm dist}\nolimits}

\def\supp{\mathop{\rm supp}\nolimits}

\newshortcut{\bzero}{\mathbf{0}}
\newshortcut{\bone}{\mathbf{1}}

\DeclareSymbolFont{symbolsC}{U}{txsyc}{m}{n}
\DeclareMathSymbol{\Perp}{\mathrel}{symbolsC}{121}

\newshortcut{\Exp}{\operatorname{Exp}}
\newshortcut{\Uniform}{\operatorname{Uniform}}
\newshortcut{\Poisson}{\operatorname{Poisson}}
\newshortcut{\Binomial}{\operatorname{Binomial}}

\newshortcut{\id}{\operatorname{id}}
\newshortcut{\tr}{\operatorname{tr}}
\newshortcut{\Tr}{\operatorname{Tr}}
\newshortcut{\rank}{\operatorname{rank}}
\newshortcut{\adj}{\operatorname{adj}}

\newshortcut{\psd}{\succeq}
\newshortcut{\pd}{\succ}

\newshortcut{\smallgap}{\vspace{1em}}

\AtBeginDocument{
\setlength\abovedisplayskip{0.5em}
\setlength\belowdisplayskip{0.5em}
}

\setlist{itemsep=0em,topsep=0em}

\newtheorem*{theorem*}{Theorem}

\newtheorem*{lemma*}{Lemma}

\newtheorem*{corollary*}{Corollary}

\newtheorem*{observation*}{Observation}

\newtheorem*{proposition*}{Proposition}

\newtheorem*{claim*}{Claim}

\theoremstyle{definition}

\newtheorem*{assumption*}{Assumption}

\newtheorem*{definition*}{Definition}

\newtheorem*{example*}{Example}

\newtheorem*{exercise*}{Exercise}

\theoremstyle{remark}

\newtheorem*{remark*}{Remark}

\makeatletter

\@addtoreset{theorem}{section}

\definecolor{mygray}{RGB}{215,215,215}
\definecolor{myblue}{RGB}{17,94,140}

\makeatletter
\tcbset{
	mytheorem/.code args={#1#2#3#4}{%
		\refstepcounter{#2}\label{#4}%
		\pgfkeysalso{title={\setbox\z@=\hbox{#1~\csname the#2\endcsname\ }\hangindent\wd\z@\hangafter=1 \mbox{#1~\csname the#2\endcsname\ }#3}}},%
}

\newcommand{\mtcbmaketheorem}[5]{%
	\newtcolorbox{#1}[3][]{#3,mytheorem={#2}{#4}{##2}{#5:##3},##1}%
}
\makeatother

\tcbset{
	simplestyle/.style={
		breakable,
		freelance,
		fonttitle=\bfseries\sffamily
	},
	thmstyle/.style={
		breakable,
		freelance,
		boxrule=2pt,
		width=\linewidth,
		frame code={%
			\path[fill=myblue,draw=myblue!75!black]
			(frame.north west) -- ([xshift=-8pt]frame.north east) --
			([yshift=-8pt]frame.north east) -- (frame.north east|-interior.north east) --
			(frame.north west|-interior.north west) -- cycle;
		},
		interior titled code={
			\path[fill=mygray!80,draw=mygray]
			(frame.west|-interior.north west) -- (frame.east|-interior.north east) --   
			([yshift=8pt]frame.east|-interior.south east) -- 
			([xshift=-8pt]frame.east|-interior.south east) --
			(frame.west|-interior.south west) -- cycle;
		},
		fonttitle=\bfseries\sffamily
	},
	defstyle/.style={
		breakable,
		freelance,
		boxrule=2pt,
		width=\linewidth,
		frame code={%
			\path[top color=myblue!50,bottom color=myblue!50,
			middle color=myblue!50]
			([xshift=8pt]frame.north west) -- ([xshift=-8pt]frame.north east) --
			([yshift=-8pt]frame.north east) -- 
			(frame.north east|-interior.north east) --
			(frame.north west|-interior.north west) -- 
			([yshift=-8pt]frame.north west) -- cycle;
		},
		interior titled code={
			\path[fill=mygray!80]
			(frame.west|-interior.north west) -| 
			([yshift=8pt]frame.east|-interior.south east) -- 
			([xshift=-8pt]frame.east|-interior.south east) -- 
			([xshift=8pt]frame.west|-interior.south west) -- 
			([yshift=8pt]frame.west|-interior.south west) -- cycle;
			\path[fill=myblue] 
			([xshift=0.5\pgflinewidth,yshift=1.5pt]frame.west|-interior.north west) 
			rectangle 
			([xshift=-0.5\pgflinewidth,yshift=-1.5pt]frame.east|-interior.north east);
		},
		fonttitle=\bfseries\sffamily\normalcolor
	}
}

\mtcbmaketheorem{problembox}{Problem}{simplestyle}{theorem}{ex}
\mtcbmaketheorem{theorembox}{Theorem}{simplestyle}{theorem}{ex}
\mtcbmaketheorem{corollarybox}{Corollary}{simplestyle}{theorem}{ex}

\usepackage{todonotes}

\begin{document}

\def\spacingset#1{\renewcommand{\baselinestretch}%
{#1}\small\normalsize} \spacingset{1}


\if0\blind
{
    \title{Efficient GPU-accelerated fitting of observational health-scaled stratified and time-varying Cox models}
    \author{Jianxiao Yang$^1$, \\Martijn J.~Schuemie$^{2,3}$, \\Marc A.~Suchard$^{1, 2, 4, 5}$}
    \date{
        $^1$Department of Computational Medicine, David Geffen School of Medicine at UCLA, Los Angeles, CA, USA\\
        $^2$Department of Biostatistics, Fielding School of Public Health at UCLA, Los Angeles, CA, USA\\
        $^3$Johnson \& Johnson, Titusville, NJ, USA\\
        $^4$Department of Human Genetics, David Geffen School of Medicine at UCLA, Los Angeles, CA, USA\\
        $^5$VA Informatics and Computing Infrastructure, US Department of Veterans Affairs, Salt Lake City, UT, USA
    }
    \maketitle
    
%
%
%
    \newpage
} \fi

\if1\blind
{
  \bigskip
  \bigskip
  \bigskip
  \begin{center}
    {\LARGE\bf Title}
\end{center}
  \medskip
} \fi

\bigskip
\begin{abstract}
The Cox proportional hazards model stands as a widely-used semi-parametric approach for survival analysis in medical research and many other fields.
Numerous extensions of the Cox model have further expanded its versatility.
Statistical computing challenges arise, however, when applying many of these extensions with the increasing complexity and volume of modern observational health datasets.
To address these challenges, we demonstrate how to employ massive parallelization through graphics processing units (GPU) to enhance the scalability of the stratified Cox model, the Cox model with time-varying covariates, and the Cox model with time-varying coefficients.
First we establish how the Cox model with time-varying coefficients can be transformed into the Cox model with time-varying covariates when using discrete time-to-event data.
We then demonstrate how to recast both of these into a stratified Cox model and identify their shared computational bottleneck that results when evaluating the now segmented partial likelihood and its gradient with respect to regression coefficients at scale.
These computations mirror a highly transformed segmented scan operation.
While this bottleneck is not an immediately obvious target for multi-core parallelization, we convert it into an un-segmented operation to leverage the efficient many-core parallel scan algorithm.
Our massively parallel implementation significantly accelerates model fitting on large-scale and high-dimensional Cox models with stratification or time-varying effect, delivering an order of magnitude speedup over traditional central processing unit-based implementations.
\end{abstract}

\noindent%
\vfill

\newpage
\spacingset{1.5} 

\section{Introduction}

The Cox proportional hazards model \citep{cox1972regression} reigns as the most popular semi-parametric approach in survival analysis, providing valuable insights into the relationships between covariates and the hazard function.
Numerous extensions of the Cox model, such as  stratification, time-varying covariates, and time-varying coefficients, have been developed to accommodate the dynamic nature of real-world research problems.
The stratified Cox model \citep{therneau2000cox} can handle covariates that do not satisfy the proportional hazards (PH) assumption by stratifying the data on them.
Additionally, Crowley and Hu \cite{crowley1977covariance} introduces a method to handle time-varying covariates in the Cox model, enabling the analysis of covariate changes over time.
Similarly, Zucker and Karr \cite{zucker1990nonparametric} extends the model to incorporate time-varying coefficients, facilitating the examination of how covariates' effects evolve over time.

While these contributions have significantly improved the versatility of the Cox model, they often encounter challenges in handling the ever-expanding size of modern observational datasets, particularly electronic health record (EHR) and administrative claims sources.
EHR and claims datasets now encompass up to hundreds of millions of individuals, involving hundreds of thousands of patient characteristics, diseases, medications, and procedures occurring over decades of patient lives \citep{hripcsak2016characterizing, hripcsak2021drawing}.
The Cox model itself exhibits quadratic growth with sample size in its na\"ive implementation, further exacerbating the computational burden.
Moreover, the various extensions of the Cox model introduce additional complexities, making the model fitting process even more challenging.
Therefore, statistical computing challenges arise both from the large-scale data and the intricacies of the extended Cox models.
Addressing these challenges calls for the utilization of advanced computing techniques to scale up survival analysis using these semi-parametric models.

Research studies on these extensions of the Cox model mainly focus on proposing the novel estimation approaches and remain limited to small to moderate-sized data \citep{tian2005cox, thackham2020maximum, he2022stratified}.
The utilization of graphics processing units (GPUs) and fine-grained parallelism to accelerate statistical computations is a relatively new and emerging area within the field of medical statistics.
For instance, Suchard et al. \cite{suchard2013massive} showcase that massive parallelization through GPUs yields one to two orders of magnitude improvement over traditional central processing unit (CPU) parallelization when applied to a computationally demanding self-controlled case series models.
Ko et al. \cite{ko2022high} investigate GPU parallelization for proximal gradient descent on modest sized $\ell_1$ regularized dense Cox regression using off-the-self software, such as PyTorch.
These studies highlight the significant performance gains that leveraging GPUs achieves in complex statistical computations.
Recently, we have proposed a time- and memory-efficient GPU implementation of regularized Cox and Fine-Gray regression models for analyzing large-scale, time-to-event data with and without competing risks \citep{yang2023massive}.

In this manuscript, we leverage massive parallelization to enhance the scalability of the seemingly less parallelizable stratified Cox model, the Cox model with time-varying covariates, and the Cox model with time-varying coefficients.
Specifically, we demonstrate that the Cox model with time-varying coefficients can be transformed into the Cox model with time-varying covariates when utilizing discrete time-to-event data.
To accomplish this, we reveal that the Cox model with time-varying covariates shares a similar partial likelihood structure as the stratified Cox model.
Consequently, all three extensions of the Cox model we investigate encounter the same computational bottleneck due to segmented scan, particularly in cases with high stratification or frequent changes in time-varying effects.
Recognizing that segmented operations are not immediately obviously parallelizable, we address this issue by transforming the computational bottleneck into un-segmented operations \citep{sengupta2008efficient}.
While even un-segmented scans, with their apparent serial output dependence, may not intuitively appear readily parallelizable, we leverage a single-pass parallel scan algorithm implemented in the cutting-edge GPU accelerated library \texttt{CUB} \citep{merrill2015cub}.
We implement our work in the easy-to-use \texttt{R} package \texttt{Cyclops}.
Our GPU implementation significantly accelerates the computation of fitting these complex models on large-scale and high-dimensional simulated and real-world data by an order of magnitude compared to a similarly optimized CPU implementation, reducing the fitting time for the analyses containing one million patients from nearly one day to just one to two hours.

\section{Methods}

\subsection{The stratified Cox proportional hazards model}\label{ch:2.1}

The stratified Cox model provides a straightforward approach to handle a covariate that does not satisfy the proportional hazards (PH) assumption.
For example, we can stratify the observations into different strata based on their disease stage when the disease stage does not meet the PH assumption, so that only the observations within each stratum share the same baseline hazard function.
Let $T_{ki}$ denote the time-to-event time and $C_{ki}$ be the right-censoring time for individual $i$ in stratum $k$, $i = 1, ..., n_k$, and $k = 1, ..., K$.
Here $n_k$ is the sample size of stratum $k$, and $K$ is the number of strata in the stratified Cox model.
Then the total sample size is $N = \sum_{k=1}^K n_k$.
For an individual, the observed time is given by $Y_{ki} = \min{(T_{ki}, C_{ki})}$, and $\delta_{ki}$ indicates whether the individual fails or is censored at $Y_{ki}$ by the value $1$ versus $0$.
Let $\mathbf{x}_{ki}$ be a $P$-dimensional covariate vector for this individual.
For this stratified Cox model with $K$ strata, the hazard for an individual from stratum $k$ is
\begin{eqnarray}
h_k(t | \mathbf{x}) = h_{0k}(t)\exp{\left(\mathbf{x}\transpose\bbeta\right)},
\end{eqnarray}
where $h_{0k}(t)$ is an unspecified baseline hazard function for stratum $k$, and $\bbeta = (\beta_1, \beta_2, \ldots, \beta_P)\transpose$ is a set of unknown, underlying model parameters that we wish to estimate.
Note that unlike the classic Cox model that assumes the same baseline hazard function $h_{0}(t)$ for all individuals, a stratified Cox model allows a distinct baseline hazard function for each stratum but a common or shared set of model parameters $\bbeta$.

The partial likelihood of the stratified Cox model falls out as the product of the partial likelihood contributions from all strata:
\begin{eqnarray} \label{strat_l}
l_{\text{\tiny partial}}(\bbeta) = \prod_{k = 1}^K\prod_{i = 1}^{n_k} \left[\frac{\exp \left(\mathbf{x}_{ki}\transpose\bbeta\right)} {\sum_{r \in R_k(Y_{ki})}\exp \left(\mathbf{x}_{kr}\transpose\bbeta\right)} \right]^{\delta_{ki}},
\end{eqnarray}
where $R_k(Y_{ki}) = \{r : Y_{kr} \geq Y_{ki}\}$ consists of the set of subjects who remain ``at risk'' for an event at time $Y_{ki}$ in stratum $k$.

When dealing with high-dimensional data, researchers may add an $\ell_1$-penalty into all or a large subset of the model parameters $\pi(\bbeta) = \sum_j\pi(\beta_j | \gamma_j) = - \sum_j\gamma_j |\beta_j|$ to the log-partial likelihood and achieve regularization through estimating the joint penalized likelihood \citep{genkin2007large, mittal2014high}.
In practice, one generally assumes $\gamma_j = \gamma$ $\forall j$ and chooses $\gamma$ through cross-validation \citep{mittal2014high}.

\subsection{The Cox model with time-varying covariates}\label{ch:2.2}

In traditional Cox regression analysis, we usually only measure the covariates at baseline once.
However, certain covariates may change during the follow-up period, such as repeated measurements in medical research.
The Cox model is able to encompass time-varying covariates using a hazard function
\begin{eqnarray}
h(t | \mathbf{x}(t)) = h_{0}(t)\exp{\left(\mathbf{x}(t)\transpose \bbeta\right)}.
\end{eqnarray}
The partial likelihood is similar in form to the classic Cox model:
\begin{eqnarray} \label{time_cov_l}
L_{\text{\tiny partial}}(\bbeta) = \prod_{i = 1}^{N}\left[\frac{\exp \left(\mathbf{x}_{i}(Y_i)\transpose\bbeta\right)}{\sum_{r \in R(Y_{i})}\exp \left(\mathbf{x}_{r}(Y_i)\transpose\bbeta\right)} \right]^{\delta_i}.
\end{eqnarray}

In practice, measurements of time are often discrete \citep{tutz2016modeling}, thus we consider the common case where $\mathbf{x}_i(t)$ can be seen as a piecewise-constant function on $K$ time intervals for individual $i = 1, 2, \ldots, N$ \citep{ngwa2016comparison} such that
\begin{align}
  \mathbf{x}_i(t) =
    \begin{cases}
      \mathbf{x}_{1i} & \text{for $t \in [t_0, t_1)$}\\
      \mathbf{x}_{2i} & \text{for $t \in [t_1, t_2)$}\\
      \ldots \\
      \mathbf{x}_{Ki} & \text{for $t \in [t_{K-1}, t_K)$}
    \end{cases}
\end{align}
for some constants $0 = t_0 < t_1 < t_2 < \cdots < t_{K} = \max(Y_1, Y_2, \ldots, Y_N)$.
Correspondingly, we can transform the time-fixed data, including the survival time and event indicator, into the discrete time intervals mentioned above.
This type of data is also referred to as discrete time-to-event data \citep{tutz2016modeling}.
Let set $S_k = \{i: Y_i \in [t_{k-1}, t_k)\}$ for $k = 1, 2, \ldots, K$.
Then the partial likelihood becomes
\begin{eqnarray} \label{time_cov_strat_l}
L_{\text{\tiny partial}}(\bbeta)
&=& \prod_{k = 1}^{K}\prod_{i \in S_k}\left[\frac{\exp \left(\mathbf{x}_{ki}\transpose\bbeta\right)}{\sum_{r \in R(Y_{i})}\exp \left(\mathbf{x}_{kr}\transpose\bbeta\right)} \right]^{\delta_i} \\
&=& \prod_{k = 1}^{K}\prod_{i = 1}^N\left[\frac{\exp \left(\mathbf{x}_{ki}\transpose\bbeta\right)}{\sum_{r \in R(Y_{ki}^{\text{\tiny (aug)}})}\exp \left(\mathbf{x}_{kr}\transpose\bbeta\right)} \right]^{\delta_{ki}^{\text{\tiny (aug)}}},
\end{eqnarray}
where augmented variables $Y_{ki}^{\text{\tiny (aug)}} = \min{(Y_i, t_k)}$ and $\delta_{ki}^{\text{\tiny (aug)}}$ indicates whether the individual $i$ fails or is censored at time $Y_{ki}^{\text{\tiny (aug)}}$.

Although observations in different time intervals still share the same baseline hazard function, the partial likelihood follows the same structure as the stratified Cox model.
Now we are able to fit a Cox model with time-varying covariates as a Cox model stratified on $K$ time intervals with the augmented design matrix $\mathbf{X}^{\text{\tiny (aug)}} = \begin{bmatrix}
\mathbf{x}_{11}
& \mathbf{x}_{12}
& \cdots
& \mathbf{x}_{1N}
& \mathbf{x}_{21}
& \cdots
& \mathbf{x}_{KN}
\end{bmatrix}
\in \mathbb{R}^{(K \times N) \times P}$, augmented observed time $\mathbf{Y}^{\text{\tiny (aug)}} \in \mathbb{R}^{(K \times N)}$, and augmented event indicator $\bm{\delta}^{\text{\tiny (aug)}} \in \mathbb{R}^{(K \times N)}$.

\subsection{The Cox model with time-varying coefficients}\label{ch:2.3}

Time-varying coefficient arises in survival analysis when a covariate's effect on the outcome is not constant over the follow-up time.
For instance, we usually assume that the COVID vaccine efficacy varies before and after 14 days of vaccination.
The proportional hazards assumption of the Cox model does not hold in this situation, as the hazard ratio comparing two specifications of a time-varying coefficient is no longer independent of time.

The extension of a Cox model with time-varying coefficients has a hazard function
\begin{eqnarray}
h(t | \mathbf{x}) = h_{0}(t)\exp{\left(\mathbf{x}\transpose \bbeta(t)\right)},
\end{eqnarray}
where $\bbeta(t) = (\beta_1(t), \beta_2(t), \ldots, \beta_P(t))\transpose$.
The partial likelihood is as follows:
\begin{eqnarray} \label{time_coef_l}
L_{\text{\tiny partial}}(\bbeta) = \prod_{i = 1}^{N}\left[\frac{\exp \left(\mathbf{x}_{i}\transpose\bbeta(Y_i)\right)}{\sum_{r \in R(Y_{i})}\exp \left(\mathbf{x}_{r}\transpose\bbeta(Y_i)\right)} \right]^{\delta_i}.
\end{eqnarray}
Often one can specify $\beta_j(t)$ as a simple step function \citep{zhang2018time}.
Without loss of generality, suppose $x_1$ has a time-varying effect on the outcome before and after time $t_s$, then $\bbeta(t)$ contains entries
\begin{align}
  \beta_j(t) =
    \begin{cases}
      \beta_{11} & \text{for $j = 1, t < t_s$}\\
      \beta_{12} & \text{for $j = 1, t \geq t_s$}\\
      \beta_j & \text{for $j = 2, 3, \ldots, P$}.
    \end{cases}
\end{align}
Then some simple calculation shows that
\begin{align}
  \mathbf{x}_i\transpose \bbeta(t) =
    \begin{cases}
      x_{i1}\beta_{11} + \sum_{j = 2}^P x_{ij}\beta_j & \text{for $t < t_s$}\\
      x_{i1}\beta_{12} + \sum_{j = 2}^P x_{ij}\beta_j& \text{for $t \geq t_s$}
    \end{cases}
    = \mathbf{x}_i^{\transpose}(t) \bbeta^{\text{\tiny (aug)}},
\end{align}
where
\begin{align}
  \mathbf{x}_i(t) =
    \begin{cases}
      (x_{i1}, 0, x_{i2}, \ldots, x_{iP})^{\transpose} & \text{for $t < t_s$}\\
      (0, x_{i1}, x_{i2}, \ldots, x_{iP})^{\transpose} & \text{for $t \geq t_s$}
    \end{cases}
   \quad\text{and}\quad
   \bbeta^{\text{\tiny (aug)}} = (\beta_{11}, \beta_{12}, \beta_2, \ldots, \beta_P)\transpose.
\end{align}
In this way, we can model a time-varying coefficient as a set of time-varying covariates and further turn a Cox model with time-varying covariates to a Cox model that stratified on time intervals.

\subsection{Maximum partial likelihood estimation using cyclic coordinate descent}\label{ch:2.4}

To maximize $L_{\text{\tiny partial}}(\bbeta)$ with or without a regularization penalty with respect to $\bbeta$, we consider a cyclic coordinate descent (CCD) algorithm which cycles through each covariate $\beta_j$ and updates it by a Newton approach while holding all other covariates as contants \citep{genkin2007large, mittal2014high}.
Specifically, when we cycle through the $j$-th covariate at $l$-th iteration, we can rewrite the partial log likelihood using a second-order Taylor approximation:
\begin{eqnarray}
g(\beta_j) \approx g(\beta_j^{(l-1)}) + g'(\beta_j^{(l-1)})(\beta_j - \beta_j^{(l-1)}) + \frac{1}{2}g''(\beta_j^{(l-1)})(\beta_j - \beta_j^{(l-1)})^2,
\end{eqnarray}
where $g'(\beta_j^{(l-1)})$ and $g''(\beta_j^{(l-1)})$ represent the one-dimensional objective gradient and Hessian with respect to $\beta_j$ evaluated at previous iteration, respectively.
To minimize the objective, the new estimate at $l$-th iteration falls out as:
\begin{eqnarray}\label{eq:newton}
\beta_j^{(l)} = \beta_j^{(l-1)} + \Delta\beta_j^{(l)} = \beta_j^{(l-1)} - \frac{g'(\beta_j^{(l-1)})}{g''(\beta_j^{(l-1)})}.
\end{eqnarray}

This CCD approach avoids the inversion of large Hassian matrices in the high-dimensional setting and only requires the scalar gradients and Hessians.
This opens up opportunities for fine-grain parallelization (discussed in the next section).
When the objective function is simply the negative log partial likelihood of stratified Cox model, the gradient and Hessian fall out as
\begin{eqnarray}\label{eq:strat_g}
g'(\beta_j)
= -\sum_{k = 1}^K \sum_{i = 1}^{n_k} x_{kij}\delta_{ki}
+ \sum_{k = 1}^K \sum_{i = 1}^{n_k}
\delta_{ki} \frac
{\sum_{r \in R_k(Y_{ki})}x_{krj}\exp \left(\mathbf{x}_{kr}^{\prime}\bbeta\right)}
{\sum_{r \in R_k(Y_{ki})}\exp \left(\mathbf{x}_{kr}^{\prime}\bbeta\right)}
\end{eqnarray}
and
\begin{eqnarray}\label{eq:strat_h}
g''(\beta_j) = \sum_{k = 1}^K \sum_{i = 1}^{n_k}\delta_{ki} \frac
{\sum_{r \in R_k(Y_{ki})}x_{krj}^2\exp \left(\mathbf{x}_{kr}^{\prime}\bbeta\right)}
{\sum_{r \in R_k(Y_{ki})}\exp \left(\mathbf{x}_{kr}^{\prime}\bbeta\right)} -
\sum_{k = 1}^K \sum_{i = 1}^{n_k}\delta_{ki} \left(\frac
{\sum_{r \in R_k(Y_{ki})}x_{krj}\exp \left(\mathbf{x}_{kr}^{\prime}\bbeta\right)}
{\sum_{r \in R_k(Y_{ki})}\exp \left(\mathbf{x}_{kr}^{\prime}\bbeta\right)}\right)^2.
\end{eqnarray}
Note that the repeated evaluations in the numerator and denominator of Equations \eqref{eq:strat_g} and \eqref{eq:strat_h} constitute the computational bottleneck.
We can conveniently add the penalty $\pi(\bbeta)$ for $\bbeta$ into the objective function when the regularization is needed.

With the step size $\Delta\beta_j^{(l)} = - \frac{g'(\beta_j^{(l-1)})}{g''(\beta_j^{(l-1)})}$ derived from Newton's method, we further improve the convergence by restricting the step size through a trust region approach \citep{genkin2007large}.
Specifically, we initialize a trust region half-width $\Delta_j^{\text{(0)}} = 1$ and update it as $\Delta_j^{(l)} = \max \{2 |\Delta \beta_j^{(l-1)}|, \Delta_j^{(l-1)} / 2 \}$.
Subsequently, we constrain the step size when updating the parameter at iteration $l$ as follows:
\begin{eqnarray}
\beta_j^{(l+1)} = \beta_j^{(l)} + \text{sgn}\left(\Delta\beta_j^{(l)}\right) \min \{ |\Delta\beta_j^{(l)}|, \Delta_j^{(l)} \}.
\end{eqnarray}

When not considering the penalty $\pi(\bbeta)$, the objective function is clearly convex and differentiable everywhere in $\bbeta$.
However, the $\ell_1$ penalty is convex but nondifferentiable at origin.
Therefore, we compute the directional derivatives \citep{wu2008coordinate} in both directions by settting $\text{sgn}\left(\Delta\beta_j^{(l)}\right) = +1$ and $\text{sgn}\left(\Delta\beta_j^{(l)}\right) = -1$ at origin.
If both directional derivatives are non-negative, we skip the update for this iteration.
If either directional derivative is negative, we update $\beta_j$ in that direction to minimize the objective.
Since the objective is convex, it is impossible for both directional derivatives to be negative.
While we do not provide a rigorous proof of convergence, the trust region method was demonstrated effectiveness in our simulations and real-world experiments.

\subsection{GPU-accelerated statistical computing strategies}\label{ch:2.5}

There are two distinct strategies in parallel computing: coarse-grained parallelism and fine-grained parallelism.
The former divides the problem into a small number of large tasks due to limitations in data communication between cores.
This strategy is typically used in parallel computing on clusters and multi-core CPUs \citep{suchard2010understanding}.
In contrast, fine-grained parallelism breaks down computational workloads into a large number of tiny tasks that run in almost lockstep.
This strategy requires significant data communication between cores and is well-suited for GPUs since they have shared memory \citep{holbrook2020massive}.

While conventionally used for graphic rendering, GPUs are growing in popularity in recent years for their potential to accelerate various scientific and engineering applications.
In this section, we briefly review the basics of parallel computing on GPUs and discuss strategies for accelerating statistical computing using fine-grained parallelism.

Understanding the hierarchical structure of threads and memory of GPUs is crucial for achieving high performance in GPU programming.
Each thread can access its own set of processor registers and local memory for thread-private variables.
Collections of up to 512 threads on current hardware group together as a thread block that has a limited shared memory only accessible to the threads within this block, enabling efficient data communication within the same block.
A grid is a collection of blocks that execute the same kernel function.
GPUs also sport large, but off-chip global memory accessible by all executing threads, regardless of if they live in the same or different blocks.
Accessing consecutive addresses in this global memory by threads in the same block leads to coalesced transactions, delivering much higher memory high-bandwidth than for most CPUs.
It is important to note that the register memory, local memory of a thread, and shared memory of a block are high-speed on-chip memory, while accessing global memory is relatively slower.
Therefore, minimizing the number of global memory transactions can greatly improve the performance of GPU programs.

GPU parallel computing works by executing kernels that are functions that run in parallel across a set of parallel threads, following a single instruction, multiple thread (SIMT) architecture.
To maximize the utilization of hardware resources, contemporary GPUs employ warp and lockstep execution \citep{nvidia2023program}.
A warp is a group of 32 parallel threads that execute the same instruction simultaneously.
If threads within a warp diverge due to data-dependent branches, the warp executes each branch path serially, and the threads converge back to the same path after all branch paths complete.
Thus, minimizing the number of diverging branches within a warp is crucial for achieving high performance.
Although a branch penalty exists when the branch divergence occurs within a warp, modern GPUs are significantly more efficient at branching code than prior parallel processors with the single-instruction, multiple-data (SIMD) architecture.

In this paper, we adopt a widely used heterogeneous computing model \citep{nvidia2023program} between the CPU and GPU for accelerating computation.
Specifically, we begin by offloading the most computationally demanding parts of the program to the GPU, while allowing the rest of the program to run on the CPU.
Note that this requires moving data between the host (CPU) and device (GPU) memory, which can be slow due to limited bandwidth between devices, thus we aim to minimize these data movements.
Once we partition the program across the host and device appropriately, the powerful computing capabilities of the GPU can make up for the expensive data movement by performing intensive calculations much faster than the CPU.

The computationally demanding portions in our case can benefit significantly more from fine-grained parallelism on GPUs than from traditional coarse-grained parallelism on CPUs.
In traditional coarse-grained parallelism on CPUs, we divide the computational work into a limited number of batches (depending on the number of available CPU cores), with each core handling the computations for a batch of samples.
In contrast, we can achieve much higher degree of parallelization on GPUs by breaking down the computation into numerous small, parallelizable tasks that require extensive inter-task communication.

\subsection{Efficient parallel segmented-scan on GPUs}\label{ch:2.6}

Here we introduce the crucial building blocks for massive sample size Cox regression analysis with and without stratification on GPUs: the \textit{scan} and \textit{segmented scan}.

A \textit{scan} procedure takes an input array $a = [a_0, a_1, \ldots, a_{n-1}]$ and an associated binary operator $\oplus$, and produces an output array $b$ where $b_i = a_0 \oplus \cdots \oplus a_i$.
Implementing a scan serially is trivial and requires $\mathcal{O}(n)$ operations.
While the output array displays obvious serial dependence in its values, this procedure remains parallelizable when communication costs between threads is low.
The na\"ive parallel scan algorithm is based on a balanced, binary tree of operations.
This algorithm reduces the computational complexity to the height of the tree $\mathcal{O}(\log_2 n)$, assuming parallel operations execute in $\mathcal{O}(1)$ at the same level on the tree.
This na\"ive algorithm, however, performs $\sum_{d = 1}^{\log_2 n}(n - 2^{d-1}) = \mathcal{O}(n \log_2 n)$ binary operations in total, even more than the sequential scan, though in parallel.
Therefore, the na\"ive algorithm is not work-efficient, as it may require additional computational resources.

A work-efficient scan algorithm arises from a \textit{reduce-then-scan} strategy that who operations visually resemble an ``hourglass'' shape consisting of an \textit{up-sweep phase} and a \textit{down-sweep phase}.
In the \textit{up-sweep phase}, we traverse the tree from leaves to root for computing a set of partial sums.
In the \textit{down-sweep phase}, we traverse back up the tree from the root to aggregate the scan output using the partial sums computed in the up-sweep phase.
Overall the two phases perform $3(n-1) = \mathcal{O}(n)$ operations and is work-efficient for large arrays.

A \textit{segmented scan} procedure takes an additional \textit{segment descriptor} with the same dimension of the input array that encodes how the input array is divided into segments.
For example, a segmented scan of the $+$ operator over an array of integers $a = [3, 1, 7, 0, 4, 1, 6, 3]$ with the segment head flags $f = [1, 0, 1, 0, 0, 1, 0, 0]$, which divide the input array into three subarrays, produces an output array $b = [3, 4, 7, 7, 11, 1, 7, 10]$.
Suppose the input array is divided into $K$ segments according to the given segment descriptor, a na\"ive parallel implementation can easily perform $K$ separate \textit{scan} procedures.
This implementation, however, can be very inefficient when $K$ is relatively large due to the overhead of launching and monitoring for the completion of $K$ separate kernels, each of which may contain only a small amount of work.
More importantly, the shorter data access patterns depending on how the input array is divided can interfere with coalescing global memory transactions.

\begin{figure}[ht]
     \centering
     \includegraphics[width=0.95\textwidth]{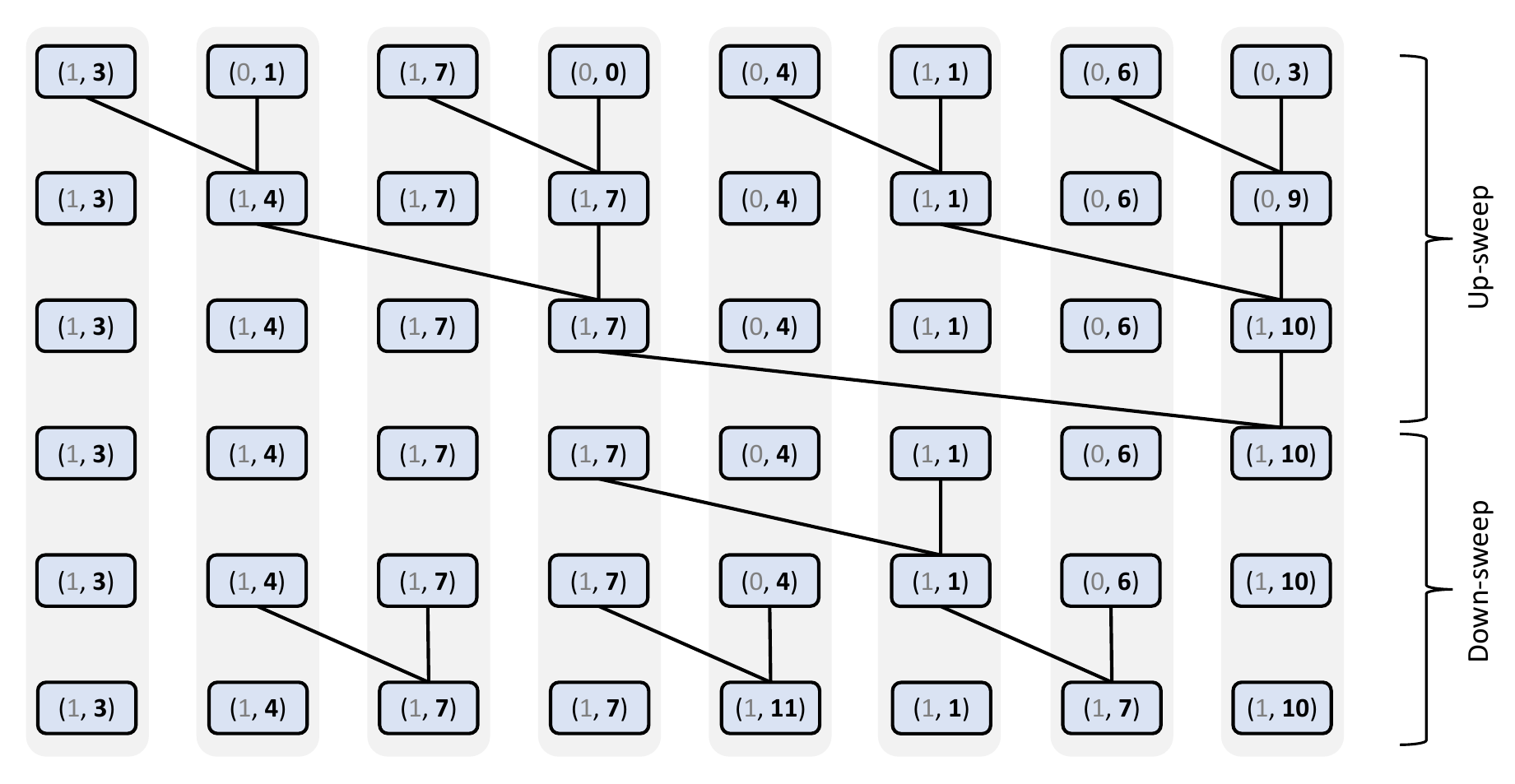}
     \caption{
     An efficient parallel \textit{segmented scan} algorithm based on binary tree.
     The efficiency of this algorithm is independent of how the input array is partitioned into segments or the number of segments.
     Each data point consists of a flag-value pair $(f_i, a_i)$, where $f_i$ is a binary indicator indicating whether the $i$-th element serves as the head of a segment, and $a_i$ represents the value intended for scan.
     The binary operator applied to these pairs is defined as $(f_i, a_i) \oplus^s (f_j, a_j) \coloneqq (f_i\ |\ f_j, \text{ if $f_j = 1$ then $a_j$ else $a_i \oplus a_j$})$, where $\oplus \coloneqq +$ in this example.
     Each vertical grey box represents an individual thread.
     Every thread executes a sequence of steps, some of which necessitate awaiting outcomes from other threads.
     The algorithm utilizes a \textit{reduce-then-scan} strategy, which can be vitalized as an ``hourglass'' shape comprising an up-sweep phase and a down-sweep phase.
     The binary tree-based parallel algorithm requires a considerable amount of inter-thread data communication, but this latency remains low in shared memory, making it suitable for GPU parallelization.}
     \label{fig:method_segmented_scan}
\end{figure}

Alternatively, we can avoid the additional overhead and suboptimal data access patterns by transforming a segmented scan into a single regular scan \citep{schwartz1980ultracomputers, blelloch1990vector, sengupta2008efficient}.
The idea is that given the input array $a$, segment head flags $f$, and associated binary operator $\oplus$, one combines $f$ and $a$ together as a new input array of flag-value pairs $(f_i, a_i)$ and constructs a new binary operator $\oplus^s$ based on $\oplus$ and $f$, such that
\begin{eqnarray}
(f_i, a_i) \oplus^s (f_j, a_j) \coloneqq (f_i\ |\ f_j, \text{ if $f_j = 1$ then $a_j$ else $a_i \oplus a_j$}).
\end{eqnarray}
Hence, the algorithm's efficiency is independent of how the input array is segmented or the total number of segments.
Figure \ref{fig:method_segmented_scan} illustrates this efficient \textit{segmented scan} algorithm, employing the \textit{reduce-then-scan} strategy.

\subsection{GPU massive parallelization for parameter estimation}\label{ch:2.7}

In this section, we aim to show how to accelerate the computational work in \ref{ch:2.4} using the parallel building block presented in \ref{ch:2.6}.

We first analyze the time complexity of our CCD algorithm presented in \ref{ch:2.4}.
In one cycle of CCD iteration, we update $\beta_j$ for $j = 1, \ldots, P$ using Newton's updates \eqref{eq:newton}, each of which in turn requires evaluating the univariate gradient \eqref{eq:strat_g} and Hessian \eqref{eq:strat_h}.
The first term in gradient \eqref{eq:strat_g} requires $\mathcal{O}(K) + \mathcal{O}(\sum_{k = 1}^K n_k = N)$ operations.
Na\"ive computation of the second term in gradient \eqref{eq:strat_g} and both terms of Hessian \eqref{eq:strat_h} require $\mathcal{O}(K) + \mathcal{O}(\sum_{k = 1}^K n_k^2)$ due to the additional inner sums in both numerator and denominator.

We can tackle fortunately all inner sums involving $R_k(Y_{ki})$ as \textit{scan}, if we arrange the observations within a stratum $k$ by their observed time $Y_{ki}$ in decreasing order.
Recall that the risk set $R_k(Y_{ki}) = \{r : Y_{kr} \geq Y_{ki}\}$ contains the  individuals in stratum $k$ who have an observed time equaling or after $Y_{ki}$, i.e.~$R_k(Y_{ki}) \in R_k(Y_{ki'} ) \forall Y_{ki} > Y_{ki'}$.
Define $S[\bnu]$ as \textit{scan} on a arbitrary vector $\bnu$.
Taking the denominator of the second term of gradient as an example, we can write the terms within stratum $k$ as
\begin{eqnarray}
\Bigg\{\sum_{r \in R_k(Y_{ki})}\exp \left(\mathbf{x}_{kr}^{\prime}\bbeta\right)\Bigg\}_{i=1}^{n_k}&
= S\left[\big\{\exp \left(\mathbf{x}_{ki}^{\prime}\bbeta\right)\big\}_{i=1}^{n_k}\right],
\end{eqnarray}
each with only cost $\mathcal{O}(n_k)$ operations.
In this way, we can reformulate the second term of gradient as $2K$ scans and reduce the time complexity from $\mathcal{O}(K) +  \mathcal{O}(\sum_{k = 1}^K n_k^2)$ to $\mathcal{O}(K) + \mathcal{O}(\sum_{k = 1}^K n_k)$.
Similarly, the time complexity of Hessian can be reduced to $\mathcal{O}(K) + \mathcal{O}(\sum_{k = 1}^K n_k)$.
However, this can still be inefficient when encountering highly-stratified data, due to the factor $K$ in the time complexity.

To this end, we further combine the data across strata $\mathbf{x}_k \in \mathbb{R}^{(n_k \times P)}$ together as $\mathbf{X} = \begin{bmatrix}
\mathbf{x}_{1}
& \cdots
& \mathbf{x}_{K}
\end{bmatrix}
\in \mathbb{R}^{N \times P}$, and turn $K$ scans to a single $K$-segmented scan.
Define head flag vectors $\mathbf{f}_k = \{f_{ki}\}_{i = 1}^{n_k}$ such that
\begin{eqnarray}
  f_{ki} =
    \begin{cases}
      1 & \text{for $i = 1$ and} \\
      0 & \text{for $i = 2, \ldots, n_k$},
    \end{cases}
\end{eqnarray}
for all $k$ and $S_{\text{\tiny seg}}[\bnu]$ as \textit{segmented scan} on the vector $\bnu$ with the head flag $\mathbf{f} = (f_{11}, \ldots, f_{1n_1}, f_{21}, \ldots, f_{Kn_K})$.  Then we can write the denominator of the second term in the gradient across all $K$ strata simply as
\begin{eqnarray}
S_{\text{\tiny seg}}\left[\big\{\exp \left(\mathbf{x}_{s}^{\prime}\bbeta\right)\big\}_{s=1}^{N}\right],
\end{eqnarray}
where $s = \sum_{k' = 1}^{k-1}{n_{k'}} + i$ for $i = 1, \ldots, n_k$ and $k = 1, \ldots, K$.
Further, define exponentiation (exp), multiplication ($\times$) and division ($/$) as element-wise operations on vectors.
The univariate gradient \eqref{eq:strat_g} and Hessian \eqref{eq:strat_h} falls out as:
\begin{eqnarray}
g'(\beta_j) &=&
- \bm{\delta}\transpose \mathbf{X}_j
+ \bm{\delta}\transpose
\frac
{S_{\text{\tiny seg}}[\mathbf{N_1}]}
{S_{\text{\tiny seg}}[\mathbf{D}]}
 \text{ and} \label{eq:parallel_g}\\
g''(\beta_j) &=&
\bm{\delta}\transpose
\frac
{S_{\text{\tiny seg}}[\mathbf{N_2}]}
{S_{\text{\tiny seg}}[\mathbf{D}]}
- \bm{\delta}\transpose
\left(
\frac
{S_{\text{\tiny seg}}[\mathbf{N_1}]}
{S_{\text{\tiny seg}}[\mathbf{D}]}
\times
\frac
{S_{\text{\tiny seg}}[\mathbf{N_1}]}
{S_{\text{\tiny seg}}[\mathbf{D}]}
\right) , \label{eq:parallel_h}
\end{eqnarray}
where
\begin{eqnarray}
\mathbf{D} &=& \exp \left(\mathbf{X}\bbeta \right), \label{eq:parallel_d}\\
\mathbf{N_1} &=& \mathbf{X}_j \times \exp \left(\mathbf{X}\bbeta \right) \text{ and} \label{eq:parallel_n1}\\
\mathbf{N_2} &=& \mathbf{X}_j \times \mathbf{X}_j \times \exp \left(\mathbf{X}\bbeta \right). \label{eq:parallel_n2}
\end{eqnarray}
Note that the vectors $\mathbf{X}_j$, $\mathbf{X}\bbeta$, $\mathbf{f}$, $\mathbf{D}$, $\mathbf{N_1}$, and $\mathbf{N_2}$ in the above equations are all of the same dimension ($N \times 1$).
Furthermore, we can avoid the costly matrix-vector multiplication $\mathbf{X}\bbeta$ in CCD by updating $[\mathbf{X}\bbeta]_s$ as shown below:
\begin{eqnarray}\label{eq:parallel_xb}
[\mathbf{X}\pmb{\beta}]_s^{\text{(new)}} = [\mathbf{X}\pmb{\beta}]_s^{\text{(old)}} + x_{sj}\Delta \beta_j,
\end{eqnarray}
for $s = 1, 2, \ldots, N$.
Regarding the vector-vector inner-products, such as $\bm{\delta}\transpose \mathbf{X}_j$, we efficiently compute these through a parallelized reduction (i.e., sum) as $\sum_{s = 1}^N \delta_s x_{sj}$. 
Note that $\mathbf{X}$ is generally sparse where most of $x_{sj}$ are zeros, resulting in relatively small computations for the above processes.
Now we have successfully reduced the time complexity of the univariate gradient \eqref{eq:strat_g} and Hessian \eqref{eq:strat_h} to $\mathcal{O}(N)$, and the time complexity of one cycle of our CCD algorithm to $\mathcal{O}(NP)$ under the stratified Cox model, regardless of the number of strata or the data distribution among them.

To parallelize the evaluation of the gradient \eqref{eq:strat_g} and Hessian \eqref{eq:strat_h} on a GPU, we generate $S = B \times IPT \times G$ threads.
Here, $B$ represents the number of concurrent threads forming a thread block, $IPT$ is the number of input items per thread, and $G = \lceil \frac{N}{B \times IPT} \rceil$ indicates the number of thread blocks.
The block size $B$ and thread grain size $IPT$ are constrained by hardware and are tunable constants.
In our implementation, we choose $B = 128$ and $IPT = 15$ for the binary-tree based kernel following the parameter settings in CUB, and $B = 256$ and $IPT = 1$ for other kernels we have developed to compute Equations \eqref{eq:parallel_d} through \eqref{eq:parallel_xb}.
Within each thread block, $B$ threads can communicate through shared memory and execute computations in parallel.
For instance, the threads first read the nonzero entries of $\mathbf{X}_j$ and $\mathbf{X}\bbeta^{\text{(old)}}$, and then update $\mathbf{X}\bbeta^{\text{(new)}}$, $\mathbf{D}$, $\mathbf{N_1}$, and $\mathbf{N_2}$ concurrently using Equations \eqref{eq:parallel_xb}, \eqref{eq:parallel_d}, \eqref{eq:parallel_n1} and \eqref{eq:parallel_n2}, respectively.
Subsequently, the threads read the values of $\mathbf{D}$, $\mathbf{N_1}$, $\mathbf{N_2}$, and the head flag $\mathbf{f}$, and perform an efficient \textit{segmented scan} operation (as detailed in \ref{ch:2.6}) using the resources of shared memory.
Finally, the threads execute the element-wise transformations and binary reductions as shown in Equations \eqref{eq:parallel_g} and \eqref{eq:parallel_h}, completing the evaluation of the gradient and Hessian for $\beta_j$ within a single iteration of the CCD process.

\section{Results}
We assess the computational efficiency of our GPU implementation versus a similar CPU implementation in fitting the extensions of Cox model to large-scale sample sizes comparable to those we often see in EHR and claims data sources.
To accomplish this, we conduct a series of synthetic experiments fitting stratified Cox models across various numbers of strata and sample sizes.
We then replicate a real-world study to evaluate the efficacy of antihypertensive drug classes using a stratified Cox model based on matching subjects through their propensity scores.
Finally, we investigate the time-varying effect of a safety outcome associated with the above drug classes employing a Cox model with time-varying coefficients.
Our computational setup comprises a system equipped with a 10-core, 3.3 GHz Intel(R) Xeon(R) W-2155 CPU, and an NVIDIA Quadro GV100 boasting 5120 CUDA cores and 32GB RAM, capable of achieving up to 7.4 Tflops double-precision floating-point performance.

\subsection{Synthetic experiments}

In this section, we illustrate the computational performance of our GPU implementation compared to the corresponding CPU implementation on stratified Cox model in the highly optimized \texttt{R} package \texttt{Cyclops} \citep{suchard2013massive,mittal2014high}.
We simulate a binary design matrix $\mathbf{X}$ under two sample sizes and two dimensions.
We randomly set $5\%$ of the entries uniformly to be $1$s to mimic the sparse pattern in the observational healthcare data.
For each sample size, we stratify the data into various number of strata.
We generate the $P$-dimensional $\bbeta$ from a standard normal distribution with mean zero and unit variance, where we set $80\%$ of the entries to $0$ to induce model parameter sparsity as well, that is
\begin{eqnarray*}
    \beta_j \sim N(0, 1) \times {\rm Bernoulli}(0.80)\ \forall\ j.
\end{eqnarray*}

\begin{figure}[H]
     \centering
     \begin{subfigure}[b]{0.8\textwidth}
         \centering
         \includegraphics[width=\textwidth]{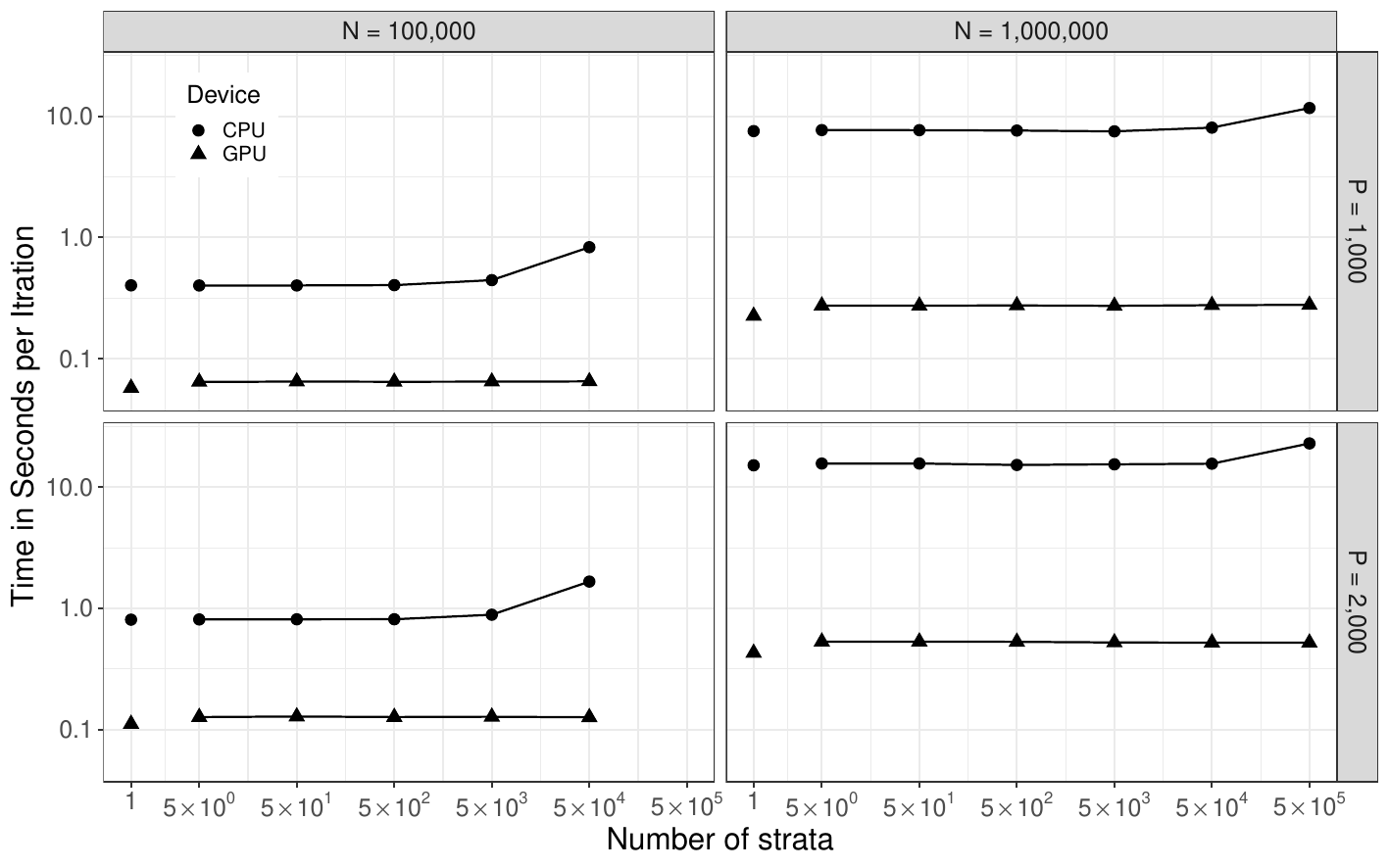}
         \label{fig:res_single_runtimes}
     \end{subfigure}
     \hfill
     \begin{subfigure}[b]{0.79\textwidth}
         \centering
         \includegraphics[width=\textwidth]{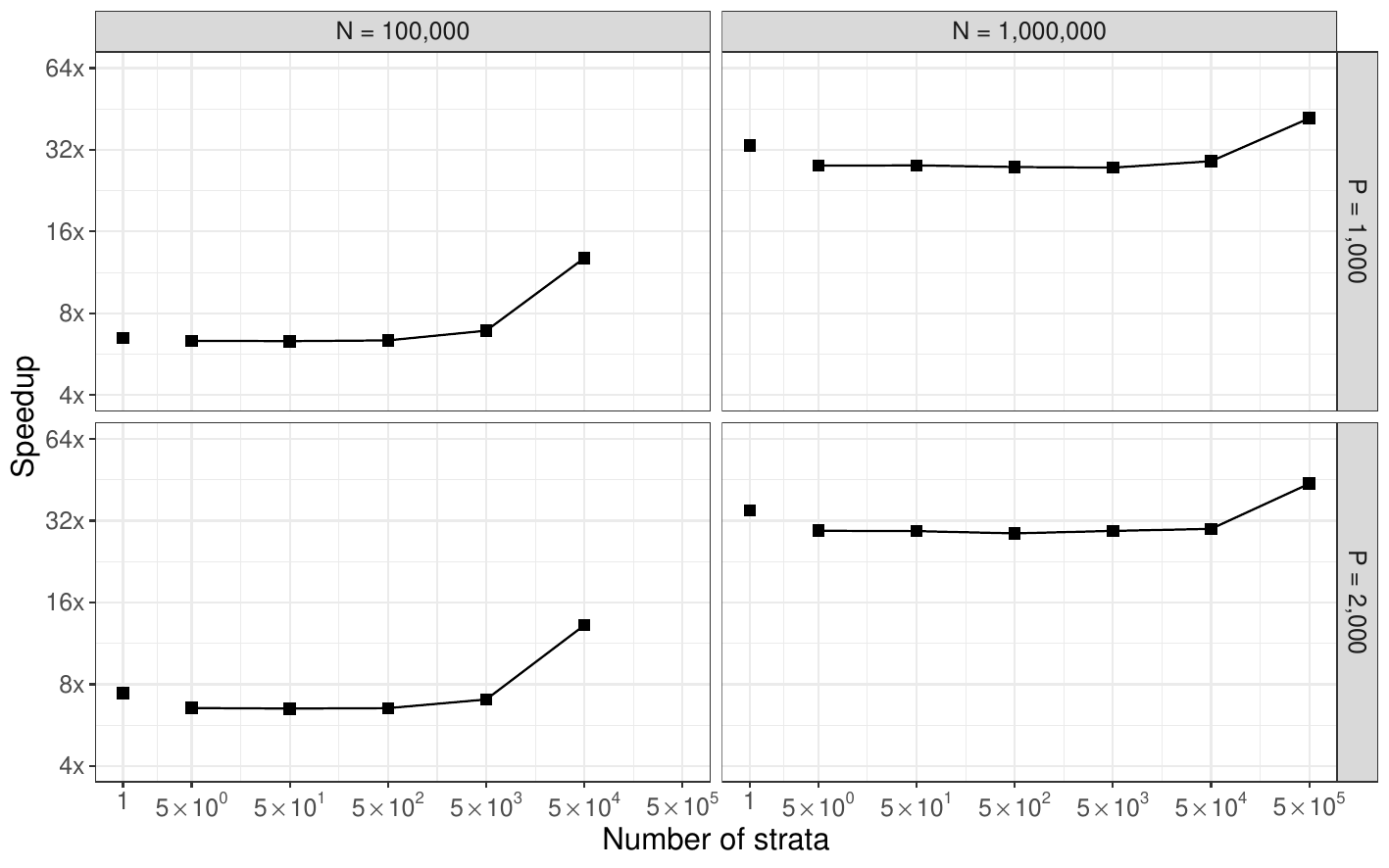}
         \label{fig:res_single_speedup}
     \end{subfigure}
        \caption{Runtimes per iteration (in seconds) and speedup of the GPU implementation relative to the CPU implementation for stratified Cox models with a fixed $\ell_1$ penalty. We conduct experiments using two sample sizes: $N = 10^5$ and $N = 10^6$, and two dimensions: $P = 1,000$ and $P = 2,000$, both with a sparsity of $95\%$. For each sample size, we fit stratified Cox models with various number of strata. In addition, we include our previous work on the unstratified Cox model for comparison purposes.}
        \label{fig:res_synthetic}
\end{figure}

We fit these simulants using a stratified Cox model under a fixed $\ell_1$ penalty with $\gamma = \sqrt{2}$, as further executing the cross-validation does alter the relative performance of the different implementations here.
Figure \ref{fig:res_synthetic} presents the runtime per iteration and the speedup of GPU paralllelization relative to a similar CPU implementation.
We also include the performance of GPU parallelization of an unstratified $\ell_1$ Cox model \citep{yang2023massive} for comparison purposes, as shown in the leftmost point in all figures.
We observe that the GPU paralllelization delivers up to a $13$-fold speedup and a $43$-fold speedup with $100,000$ samples and one million samples, respectively.
To put it into an absolute scale, we reduce the total fitting time of a stratified Cox model with one million samples and $50,000$ strata from $17$ minutes to $23$ seconds.
We also observe a rapid increase of runtimes on the CPU with the most highly stratified data, while the GPU approach consistent performs well across slightly and highly stratified data.
At $K = N / 2$, speculative instruction execution, namely poor branch prediction of where strata start, takes it toll on CPU performance.
The relevant performance improvements align with the computational complexity analysis discussed in \ref{ch:2.7}.

\subsection{Cardiovascular effectiveness of antihypertensive drug classes}

In this section, we explore the relative effectiveness of two major hypertension drug classes to demonstrate the advantages of massive parallelization within a real-world example.
While most treatment recommendations derive from randomized clinical trials that offer limited comparisons between a few agents, large-scale observational studies can provide valuable insights for estimating the relative risk of important cardiovascular and safety outcomes associated with different drug classes.
We primarily focus on two major hypertension drug classes, angiotensin-converting enzyme inhibitors (ACEIs) and thiazide or thiazide-like diuretics (THZs), and one important cardiovascular outcome, hospitalization for heart failure.
We follow a comparative new-user cohort design, as outlined in the Large-scale Evidence Generation and Evaluation across a Network of Databases for Hypertension (LEGEND-HTN) study \citep{suchard2019comprehensive}.

For our experiments here, we use patient health records on antihypertensive drug classes from the Optum\textsuperscript{\tiny\textregistered} de-identified Electronic Health Record dataset (Optum EHR).
This dataset encompasses information from 85 million individuals in the United States who are commercially or Medicare insured.
We extract a subsample of $946,911$ patients diagnosed with hypertension.
Among these individuals, $77\%$ initiate treatment with an ACEI, while the remaining $23\%$ initiate treatment with a THZ.
We consider the relative effectiveness of THZ and ACEI in preventing hospitalization for heart failure as main treatment covariate.
Additionally, we include $9,976$ baseline patient characteristic covariates, encompassing clinical condition, drug exposure, and medical procedure.
The patient characteristic covariates exhibit an average sparsity of $97\%$, indicating that only $3\%$ of the entries contain non-zero values.
We construct a propensity score model incorporating all baseline covariates \citep{tian2018evaluating} and stratify the individuals into varying number of equally-sized strata based on their propensity score estimates.
Specifically, we consider three commonly used strata configurations: $S = 5, 10,$ and $20$.
Finally, we apply the stratified Cox proportional hazards model to estimate the hazard ratio (HR) between THZ and ACEI initiation with respect to the risk of hospitalization for heart failure.
We include all patient characteristics and treatment covariates in the stratified Cox model with $\ell_1$ regularization on all covariates except the treatment covariate, and employ a 10-fold cross-validation to search for optimal tuning parameters.
Considering that statistical inference under $\ell_1$ regularization remains challenging, we calculate $95\%$ bootstrapped percentile intervals (BPIs) from bootstrap samples.
Table \ref{tb:res_stratified_LEGEND} reports the runtimes (in hours) for both our GPU parallelization and a similar CPU implementation and the HRs estimates with their BPIs.
Our GPU parallelization delivers an $11$-fold speedup across varying numbers of strata.

\begin{table}[ht]
\centering
\begin{tabular}{rrrc}
\hline
  & \multicolumn{2}{c}{Runtime (hr)} \\
\cline{2-3}
Strata & GPU & CPU & HR (95$\%$ [BPI]) \\
\hline
5  & 1.41 & 16.2 & 0.83 (0.72, 0.93) \\
10 & 1.38 & 15.8 & 0.81 (0.73, 0.90) \\
20 & 1.57 & 18.0 & 0.82 (0.73, 0.93) \\
\hline
\end{tabular}
\caption{
Computation times (in hours) for fitting the stratified Cox model for the optimum value of $\ell_1$ regularization parameter (found through 10-fold cross-validation) across GPU and CPU implementations and the HR estimates and their $95\%$ BPIs comparing the relative risk of hospitalization for heart failure risk between new-users of thiazide or thiazide-like duretics and angiotensin-converting enzyme inhibitors.
The data contains $946,911$ individuals with $9,977$ covariates.
We stratify the individuals into varying number of equally-sized strata based on propensity score estimates.}
\label{tb:res_stratified_LEGEND}
\end{table}

\subsection{Time-varying effect of safety outcome of antihypertensive drug classes}

\begin{figure}[h]
	\centering
	\subfloat{
	\begin{tikzpicture}
		\node[] (fig){
		\includegraphics[width=0.8\textwidth, keepaspectratio]{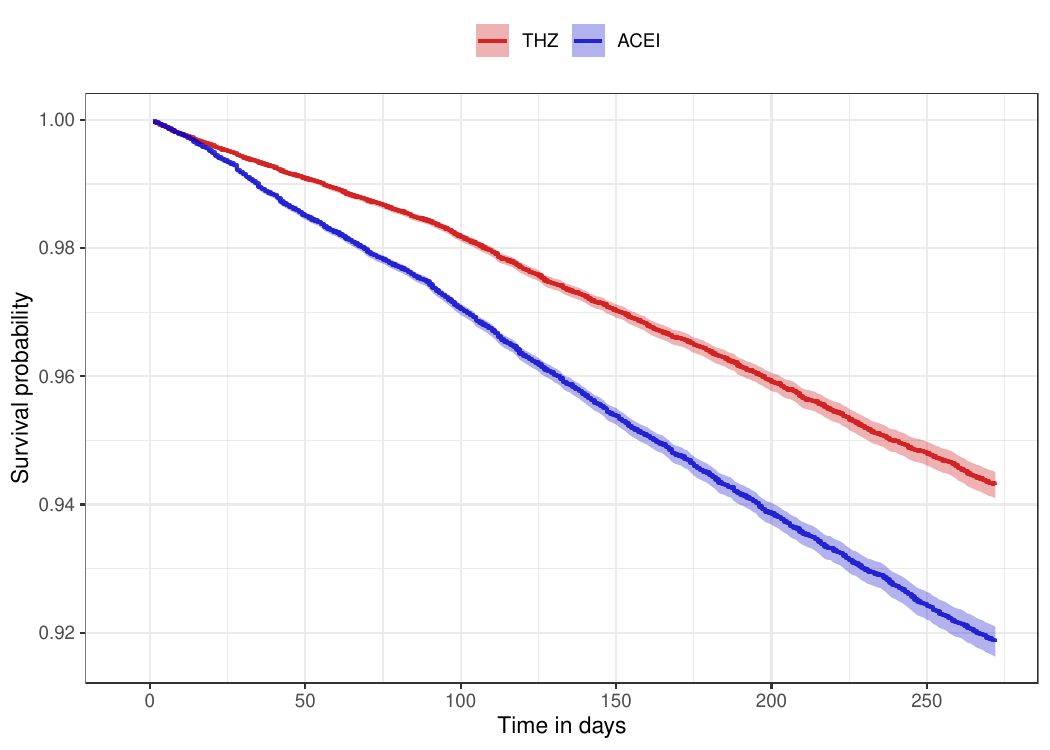}};
		\node[rectangle,
			  draw,
			  minimum width = 1.8cm,
			  minimum height = 1.2cm] (r) at (-4.1,2.9) {};
	\end{tikzpicture}
	}
	\hfill
	\subfloat{
	\includegraphics[width=0.8\textwidth]{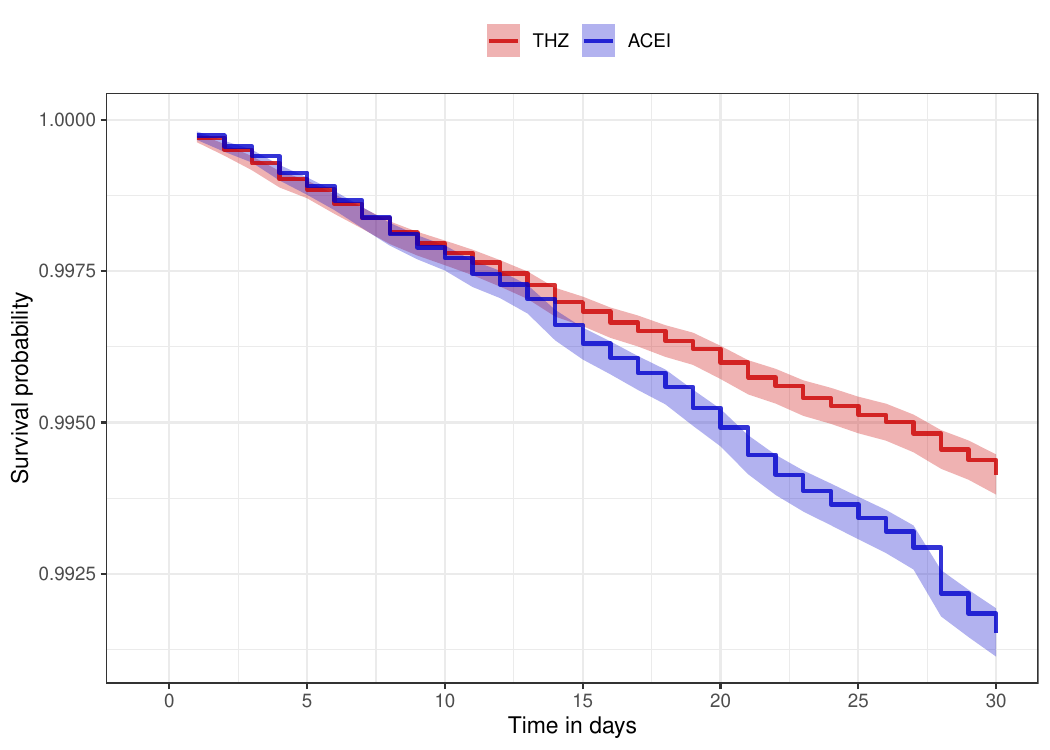}
	}
	\caption{Kaplan Meier plots showing survival of patients with cough over time. The red line represents THZ exposure, while the blue line represents ACEI exposure. The upper plot displays all the data included in the analysis, while the lower plot focuses on data within 30 days of exposure. We can see that the risks of developing cough within the first 10 days are similar in both groups, while ACEI carries a higher risk compared to THZ after 10 days.}
	\label{fig:res_cough}
\end{figure}

In this section, we investigate the time-varying effect of a safety outcome of ACEIs: cough \citep{dicpinigaitis2006angiotensin}.
We utilize a similar comparative new-user cohort design and the same dataset as in the previous section to conduct this analysis.
Rather than stratifying individuals using propensity scores, we employ a 1:1 matching strategy for THZ and ACEI new-users.
After propensity score matching, we retain a total of $407,828$ patients who developed cough with $9,666$ baseline covariates for further analysis.
The Kaplan-Meier plots in Figure \ref{fig:res_cough} shows the survival of patients with cough over time.
We can see that the relative risks of THZ and ACEI exposure on cough are different within 10 days of initiating treatment and after 10 days.
Therefore, we treat the treatment covariate as two time-varying covariates accordingly.
To assess the time-varying effects, we transform the Cox model with a time-varying coefficient to a stratified Cox model with two strata using the method explained in \ref{ch:2.3} and \ref{ch:2.2}.
The stratified Cox model for cough contains $812,432$ observations and $9,668$ covariates after appropriate data wrangling.
We apply an $\ell_1$ penalty on all covariates except the treatment covariates (within 10 days of initiating treatment and after 10 days).
We then performed a 10-fold cross-validation to identify the optimal tuning parameters.
Our GPU parallelization significantly reduces the analysis time from $21.6$ hours to $1.86$ hours.

Through massive parallelization, we find that initializing with a THZ has less risk of developing cough than initializing with an ACEI after 10 days (HR $0.67$, $95\%$ BPI $0.65 - 0.69$), while both medications exhibit similar risks within 10 days (HR $0.98$, $95\%$ BPI $0.87 - 1.13$).
Note that the HRs correspond to the exponentiated $\beta$ coefficients.
The difference between two coefficients is $0.38$ with a $95\%$ BPIs ranging from $0.30$ to $0.48$, reinforcing that the relative risks of developing cough differ within and after 10 days of initialization.
Previously, dealing with time-varying coefficients on this scale posed challenges in the original LEGEND-HTN study due to computational time burdens.
However, with the implementation of our GPU parallelization, this complex analysis has now become feasible.

\section{Discussion}

This article presents an efficient GPU implementation of stratified Cox and the time-varying Cox models for analyzing large-scale time-to-event data investigating time-varying effects, building upon our prior work on standard Cox models \citep{yang2023massive}.
We implement this efficient method in the open-source \texttt{R} package \texttt{Cyclops} \citep{suchard2013massive}.
In simulation studies, the GPU implementation for the stratified Cox model shows a speedup of $43$ times compared to the equivalent CPU implementation, even with up to 1 million samples.
In our real-world examples examining the time-varying risk of developing cough, massive parallelization significantly reduces the total runtimes from nearly a day to less than 2 hours.

The main idea of this article is to simplify a complex model into a more concise one and enhance its implementation efficiency through the innovative use of massive parallelization.
In particular, we demonstrate that a Cox model with time-varying coefficients can be transformed into a Cox model with time-varying covariates when using discrete time-to-event data.
Moreover, we identify that the Cox model with time-varying covariates shares the similar partial likelihood structure as the stratified Cox model.
Finally, we apply an efficient segmented scan algorithm to address the same computational bottleneck of the three extended models due to the similar partial likelihood structure.
This algorithm significantly accelerates likelihood, gradient, and Hessian evaluations, thereby improving the overall efficiency of our approach.

There is potential for improvement in our approach.
First, both transformations detailed in \ref{ch:2.3} and \ref{ch:2.2} require augmenting the original design matrix due to repeating the time-independent covariates over time or creating additional time-varying covariates to estimate the time-varying effect in multiple time intervals.
While data augmentation can be memory-inefficient, it is possible to save memory by developing mappings on the original data, as both data augmentation techniques involve duplications of the original data.
Additionally, the cyclic coordinate descent algorithm we use in the case with  $\ell_1$ regularization may lack of rigorous theoretical proof of convergence.
Although it is possible that the results to be oscillate around the global minimum, we have not observed this issue in our experiments.

Nonetheless, this work provides valuable tools for massively sized Cox models with stratification and time-varying effects.
In recent years, the growing interest in leveraging large-scale observational healthcare data sources has driven the demand for such models.
For instance, Shoaibi et al. \cite{shoaibi2021comparative} explored the comparative effectiveness of famotidine in hospitalized COVID-19 patients using data from the COVID-19 Premier Hospital Database, which encompasses approximately $700$ hospitals throughout the United States.
Similarly, Kim et al. \cite{kim2020comparative} studied the comparative safety and effectiveness of two popular anti-osteoporosis medications on $324,049$ patients across three electronic medical records and six claims databases.
The ability to rapidly perform analyses using such large models has opened doors to comprehensive sensitivity assessments regarding stratification designs.
Moreover, this work initiates large-scale comparative effectiveness and safety studies with time-varying effects, addressing the limitation of tools for the time-varying Cox models.

\section*{Acknowledgments}
This work was partially supported by US National Institutes of Health grants R01 HG006139, U19 AI135995, R01 AI153044 and R01 HL169954 and an Intergovernmental Personnel Act agreement with the US Department of Veterans Affairs.

\section*{Disclosures}
MJS is an employee and share-holder of Johnson \& Johnson.
The remaining authors report there are no competing interests to declare.

\bigskip
\begin{center}
{\large\bf SUPPLEMENTARY MATERIALS}
\end{center}

\begin{description}

\item[Cyclops:] R-package Cyclops containing code to perform the analysis described in the article is available at \url{https://github.com/OHDSI/Cyclops/tree/time_varying}.

\item[R scripts for executing the experiments:]
R scripts for executing the experiments in Section~ 3. are available at \url{https://github.com/suchard-group/strat_time_manuscript_supplement}.

\end{description}

\bigskip
\begin{center}
{\large\bf SUPPLEMENTARY MATERIALS}
\end{center}






\bibliographystyle{ieeetr}
\bibliography{reference.bib}

\begin{thebibliography}{10}

\bibitem{cox1972regression}
D.~R. Cox, ``Regression models and life-tables,'' {\em Journal of the Royal
  Statistical Society: Series B (Methodological)}, vol.~34, no.~2,
  pp.~187--202, 1972.

\bibitem{therneau2000cox}
T.~M. Therneau, P.~M. Grambsch, T.~M. Therneau, and P.~M. Grambsch, {\em The
  Cox Model}.
\newblock Springer, 2000.

\bibitem{crowley1977covariance}
J.~Crowley and M.~Hu, ``Covariance analysis of heart transplant survival
  data,'' {\em Journal of the American Statistical Association}, vol.~72,
  no.~357, pp.~27--36, 1977.

\bibitem{zucker1990nonparametric}
D.~M. Zucker and A.~F. Karr, ``Nonparametric survival analysis with
  time-dependent covariate effects: a penalized partial likelihood approach,''
  {\em The Annals of Statistics}, vol.~18, no.~1, pp.~329--353, 1990.

\bibitem{hripcsak2016characterizing}
G.~Hripcsak, P.~B. Ryan, J.~D. Duke, N.~H. Shah, R.~W. Park, V.~Huser, M.~A.
  Suchard, M.~J. Schuemie, F.~J. DeFalco, A.~Perotte, {\em et~al.},
  ``Characterizing treatment pathways at scale using the {OHDSI} network,''
  {\em Proceedings of the National Academy of Sciences}, vol.~113, no.~27,
  pp.~7329--7336, 2016.

\bibitem{hripcsak2021drawing}
G.~Hripcsak, M.~J. Schuemie, D.~Madigan, P.~B. Ryan, and M.~A. Suchard,
  ``Drawing reproducible conclusions from observational clinical data with
  {OHDSI},'' {\em Yearbook of Medical Informatics}, vol.~30, no.~01,
  pp.~283--289, 2021.

\bibitem{tian2005cox}
L.~Tian, D.~Zucker, and L.~Wei, ``On the {C}ox model with time-varying
  regression coefficients,'' {\em Journal of the American Statistical
  Association}, vol.~100, no.~469, pp.~172--183, 2005.

\bibitem{thackham2020maximum}
M.~Thackham and J.~Ma, ``On maximum likelihood estimation of the
  semi-parametric {C}ox model with time-varying covariates,'' {\em Journal of
  Applied Statistics}, vol.~47, no.~9, pp.~1511--1528, 2020.

\bibitem{he2022stratified}
K.~He, J.~Zhu, J.~Kang, and Y.~Li, ``Stratified {C}ox models with time-varying
  effects for national kidney transplant patients: A new blockwise steepest
  ascent method,'' {\em Biometrics}, vol.~78, no.~3, pp.~1221--1232, 2022.

\bibitem{suchard2013massive}
M.~A. Suchard, S.~E. Simpson, I.~Zorych, P.~Ryan, and D.~Madigan, ``Massive
  parallelization of serial inference algorithms for a complex generalized
  linear model,'' {\em ACM Transactions on Modeling and Computer Simulation
  (TOMACS)}, vol.~23, no.~1, p.~10, 2013.

\bibitem{ko2022high}
S.~Ko, H.~Zhou, J.~J. Zhou, and J.-H. Won, ``High-performance statistical
  computing in the computing environments of the 2020s,'' {\em Statistical
  Science}, vol.~37, no.~4, pp.~494--518, 2022.

\bibitem{yang2023massive}
J.~Yang, M.~J. Schuemie, X.~Ji, and M.~A. Suchard, ``Massive parallelization of
  massive sample-size survival analysis,'' {\em Journal of Computational and
  Graphical Statistics}, no.~just-accepted, pp.~1--23, 2023.

\bibitem{sengupta2008efficient}
S.~Sengupta, M.~Harris, M.~Garland, {\em et~al.}, ``Efficient parallel scan
  algorithms for {GPU}s,'' {\em NVIDIA, Santa Clara, CA, Tech. Rep.
  NVR-2008-003}, vol.~1, no.~1, pp.~1--17, 2008.

\bibitem{merrill2015cub}
D.~Merrill, ``Cub,'' {\em NVIDIA Research}, 2015.

\bibitem{genkin2007large}
A.~Genkin, D.~D. Lewis, and D.~Madigan, ``Large-scale {B}ayesian logistic
  regression for text categorization,'' {\em Technometrics}, vol.~49, no.~3,
  pp.~291--304, 2007.

\bibitem{mittal2014high}
S.~Mittal, D.~Madigan, R.~S. Burd, and M.~A. Suchard, ``High-dimensional,
  massive sample-size {C}ox proportional hazards regression for survival
  analysis,'' {\em Biostatistics}, vol.~15, no.~2, pp.~207--221, 2014.

\bibitem{tutz2016modeling}
G.~Tutz, M.~Schmid, {\em et~al.}, {\em Modeling Discrete Time-to-event Data}.
\newblock Springer, 2016.

\bibitem{ngwa2016comparison}
J.~S. Ngwa, H.~J. Cabral, D.~M. Cheng, M.~J. Pencina, D.~R. Gagnon, M.~P.
  LaValley, and L.~A. Cupples, ``A comparison of time dependent {C}ox
  regression, pooled logistic regression and cross sectional pooling with
  simulations and an application to the framingham heart study,'' {\em BMC
  Medical Research Methodology}, vol.~16, pp.~1--12, 2016.

\bibitem{zhang2018time}
Z.~Zhang, J.~Reinikainen, K.~A. Adeleke, M.~E. Pieterse, and C.~G.
  Groothuis-Oudshoorn, ``Time-varying covariates and coefficients in {C}ox
  regression models,'' {\em Annals of Translational Medicine}, vol.~6, no.~7,
  2018.

\bibitem{wu2008coordinate}
T.~T. Wu, K.~Lange, {\em et~al.}, ``Coordinate descent algorithms for lasso
  penalized regression,'' {\em The Annals of Applied Statistics}, vol.~2,
  no.~1, pp.~224--244, 2008.

\bibitem{suchard2010understanding}
M.~A. Suchard, Q.~Wang, C.~Chan, J.~Frelinger, A.~Cron, and M.~West,
  ``Understanding {GPU} programming for statistical computation: Studies in
  massively parallel massive mixtures,'' {\em Journal of Computational and
  Graphical Statistics}, vol.~19, no.~2, pp.~419--438, 2010.

\bibitem{holbrook2020massive}
A.~J. Holbrook, P.~Lemey, G.~Baele, S.~Dellicour, D.~Brockmann, A.~Rambaut, and
  M.~A. Suchard, ``Massive parallelization boosts big {B}ayesian
  multidimensional scaling,'' {\em Journal of Computational and Graphical
  Statistics}, vol.~30, no.~1, pp.~11--24, 2021.

\bibitem{nvidia2023program}
{NVIDIA}, ``{CUDA} {C}++ programming guide.''
  \url{https://docs.nvidia.com/cuda/cuda-c-programming-guide/index.html}, 2023.

\bibitem{schwartz1980ultracomputers}
J.~T. Schwartz, ``Ultracomputers,'' {\em ACM Transactions on Programming
  Languages and Systems (TOPLAS)}, vol.~2, no.~4, pp.~484--521, 1980.

\bibitem{blelloch1990vector}
G.~E. Blelloch, {\em Vector Models for Data-Parallel Computing}.
\newblock Cambridge: MIT press, 1990.

\bibitem{suchard2019comprehensive}
M.~A. Suchard, M.~J. Schuemie, H.~M. Krumholz, S.~C. You, R.~Chen, N.~Pratt,
  C.~G. Reich, J.~Duke, D.~Madigan, G.~Hripcsak, {\em et~al.}, ``Comprehensive
  comparative effectiveness and safety of first-line antihypertensive drug
  classes: a systematic, multinational, large-scale analysis,'' {\em The
  Lancet}, vol.~394, no.~10211, pp.~1816--1826, 2019.

\bibitem{tian2018evaluating}
Y.~Tian, M.~J. Schuemie, and M.~A. Suchard, ``Evaluating large-scale propensity
  score performance through real-world and synthetic data experiments,'' {\em
  International Journal of Epidemiology}, vol.~47, no.~6, pp.~2005--2014, 2018.

\bibitem{dicpinigaitis2006angiotensin}
P.~V. Dicpinigaitis, ``Angiotensin-converting enzyme inhibitor-induced cough:
  {ACCP} evidence-based clinical practice guidelines,'' {\em Chest}, vol.~129,
  no.~1, pp.~169S--173S, 2006.

\bibitem{shoaibi2021comparative}
A.~Shoaibi, S.~P. Fortin, R.~Weinstein, J.~A. Berlin, and P.~Ryan,
  ``Comparative effectiveness of famotidine in hospitalized {COVID}-19
  patients,'' {\em The American College of Gastroenterology}, vol.~116, no.~4,
  pp.~692--699, 2021.

\bibitem{kim2020comparative}
Y.~Kim, Y.~Tian, J.~Yang, V.~Huser, P.~Jin, C.~G. Lambert, H.~Park, S.~C. You,
  R.~W. Park, P.~R. Rijnbeek, {\em et~al.}, ``Comparative safety and
  effectiveness of alendronate versus raloxifene in women with osteoporosis,''
  {\em Scientific Reports}, vol.~10, no.~1, p.~11115, 2020.

\end{thebibliography}
\end{document}